\documentclass[12pt]{aa}
\usepackage[english]{babel}
\usepackage{graphicx}

\onecolumn
\begin{document}

\title{To the problem of spectral mimicry of supergiants}
\author{V.G. Klochkova and E.L. Chentsov}
\institute{Special Astrophysical Observatory RAS, Nizhnij Arkhyz, 369167 Russia}

\date{\today} 

\abstract{The phenomenon of spectral mimicry refers to the fact that hypergiants and post-AGB
supergiants -- stars of different masses in fundamentally different stages of their evolution 
have similar optical spectra, and also share certain other characteristics (unstable and  extended 
atmospheres, expanding gas-dust envelopes, high IR excesses). As a consequence, it is not 
always possible to distinguish post--AGB stars from hypergiants based on individual spectral 
observations in the optical range. Examples of spectral mimicry are presentes using uniform, high-quality 
spectral material obtained on the 6-m telescope of the Special Astrophysical Observatory in the 
course of long-term monitoring of high-luminosity stars. It is shown that unambiguously resolving 
the mimicry problem for individual stars requires the determination of a whole set of parameters: 
luminosity, wind parameters, spectral energy distribution, spectral features,
velocity field in the atmosphere and circumstellar medium, behavior of the parameters with time, and 
the abundances of chemical elements in the atmosphere.
 }

\authorrunning{\it Klochkova \& Chentsov}
\titlerunning{\it Spectral mimicry of supergiants}

\maketitle

\section{Introduction}

Numerous massive Population I supergiants have very well defined, similar spectra that can be used
to identify them fairly certainly. The spectra of the small number of known hypergiants are peculiar, 
with this peculiarity varying widely from object to object.
The spectra of low-mass post-AGB supergiants differ appreciably from the spectra of massive Population I
supergiants, but their spectra are often difficult to distinguish from those of hypergiants. Thus, given 
the uncertainties in the estimated distances for Galactic objects, an object with a powerful wind but with 
a lower mass and luminosity can be mistaken for a hypergiant based on its optical spectrum.
It seems paradoxical, but high-luminosity stars (hereafter -- HLSs), which are primarily distinguished 
by their masses (the most massive stars with initial masses exceeding  $20\div40\mathcal{M}_\odot$,  and 
intermediate-mass star with initial masses of  $3\div9\mathcal{M}_\odot$) and evolutionary stages, have similar 
observational properties: the properties of the optical and radio spectra, high IR excess,
and complex and time-variable velocity field, which testify to instability of their extended atmospheres
and expanding gas-dust envelopes. This intersection of observed properties for two types of objects
leads to the problem of spectral mimicry. Stars in the groups indicated above, at fundamentally 
different stages of their evolution, display similar optical
spectra and other properties (unstable and extended atmospheres, expanding gas-dust envelopes, high IR
excesses). As a rule, according to the MK classification for post-AGB stars, these are high-luminosity
objects (luminosity classes Ia, Ib, and II and absolute magnitudes M$_{\rm V}$ from ${\rm -5^m}$ to ${-7^m}$). 
However, this actually only indicates a low value of $\log g$ in the photosphere (or pseudo-photosphere) 
of the star.

For many years, we have studied stars with various masses with high mass-loss rates in their earlier
and current stages of evolution. The fundamental reasons for mass loss and its evolutionary variations
are not fully understood, underscoring the importance of determining the main characteristics of evolved
stars and studying the structures of their extended envelopes and the parameters of their stellar winds.
Our program includes studies of evolved stars with various masses: luminous blue variables (LBVs),
which are very massive evolved stars near the Eddington stability limit; B[e] stars, which are likely
intermediate-mass binary systems soon after a stage of rapid mass transfer;  white and yellow hypergiants; 
and low-mass post-asymptotic giant branch (post-AGB) supergiants with high IR excesses. The observed 
brightnesses of stars with circumstellar gas-dust envelopes in the visible is determined to an 
appreciable extent by the power of the envelope, which is related to the mass-loss rate.
A necessary aspect of our studies, which is also the most labor intensive, is determining the 
luminosities and masses of the stars, in order to fix their evolutionary stage. When studying 
individual stars, we have observed similar, and even virtually identical, spectral features in 
different objects of the types indicated above.

In Section~2, we consider some aspects of the problem of spectral mimicry in more detail, using the
properties of the objects listed in the table as examples. In Section~3, we suggest a means of resolving
this problem. The vast majority of our studies were conducted using spectral material obtained using the
same NES spectrograph [1, 2] of the 6--m telescope of the Special Astrophysical Observatory (SAO) 
applying a single method for the spectral reduction. The uniformity of this observing material has enabled us
to correctly compare features in the spectra of these different types of stars.

\begin{table}[ht!]
\caption{Main data about the various types of supergiants considered. Spectral classes 
      for a number of stars were taken from the SIMBAD database}
\medskip                                                             
\begin{tabular}{l  l  l  l  l}
\hline
 Star  &IR source \hspace{1.0cm} &   Sp & Teff, $\log g$ &Note \\
\hline
\multicolumn{5}{l}{\underline{\small Hypergiants}} \\
 6\,Cas       & IRAS\,23463+6156   &  A2.5\,Ia-0 & & [8] \\ 
 HD\,33579    & IRAS\,05057$-$6756 &  A3\,IaO &8129\,K 0.7 & [9]    \\
 V1302\,Aql   &IRC+10420           &  A2Ia$^+$   & 8500\,K, 1.0 & [10] \\
 HD\,190603   &IRAS\,20026+3204    & B5\,Ia$^+$  & 18100\,K, 2.41 &[11] \\ 
Cyg\,OB2 No.12&IRAS\,20308+4104    &  B5\,Ia$^+$ &       & [12] \\  
\multicolumn{5}{l}{\underline{\small Supergiants}} \\ 
HD\,20902 ($\alpha$\,Per) & IRAS\,03207+4941& F5\,Ib & 6579\,K 1.56   & [13] \\ 
HD\,74180     &                    &  Fo\,Ia     &7839\,K, 2.11 & [13]  \\ 
HD\,102878    &                    &  A2\,Iab    &  & SIMBAD \\ 
HD\,148379    &                    & B2\,Iab     &  & SIMBAD \\ 
HD\,168571    &                    &  BI\,Ib     & &SIMBAD \\  
\multicolumn{5}{l}{\underline{\small post-AGB supergiants}} \\
GSC\,04501$-$00166&IRAS\,01005+7910& B1.5\,Ib    &  21500\,K, 3.0& [14, 15]\\
BD+48$\degr$1220&IRAS\,05040+4820  &             & 7900\,K, 0.0  & [16] \\ 
V510\,Pup     &IRAS\,08005$-$2356  & F5\,Ib-Iab  & & [17] \\ 
V2324\,Cyg    &IRAS\,20572+4919    & F0\,III     & 7500\,K, 2.0 & [18] \\
\hline
\end{tabular}                                   
\label{stars}
\end{table}

\section{Examples of spectral mimicry} 

Let us briefly consider the main observational properties of two groups of HLSs: hypergiants and
post-AGB stars which illustrate the problem of spectral mimicry. Hypergiants are extremely 
luminous objects. They are the descendents of high-mass stars with $\mathcal{M}  \ge 20\mathcal{M}_\odot$, 
and are observed during a short-lived stage in their evolution, making then very rare objects. 
According to the generally accepted view (see, e.g., the evolutionary scenarios of Conti~[3]), 
these stars are observed as various types of HLSs during their evolution from the main sequence: 
white and yellow hypergiants, LBVs, B[e] stars, Wolf-Rayet stars. These very high-luminosity 
stars evolve, and lose matter at high rates, reaching $10^{-3}\mathcal{M}_\odot/$yr in individual 
episodes. For example, according to [4], the mass-loss rate of HD\,33579 is $10^{-5.7}\mathcal{M}_\odot/$yr, 
and the mass-loss rate of V1302\,Aql is $10^{-4.85}\mathcal{M}_\odot/$yr.
The star HD\,190603, with a similar luminosity, loses mass at the rate $10^{-5.7}\mathcal{M}_\odot/$yr~[5].

A determining factor in the evolution of these massive stars is their stellar winds, which create 
circumstellar gas-dust envelopes, which often have high densities and complex morphologies. The complex
morphology of the circumstellar structures of evolved massive stars follow from radio spectroscopy 
data obtained in the bands of various molecules and masers (see, e.g., [6, 7]).

In the course of their evolution, these stars supply the interstellar medium with matter freshly processed
by nuclear reactions; they also influence the evolution of their host galaxies via their ionizing radiation 
and mechanical energy. White hypergiants are late B and  early A stars with luminosity class Ia-0 
(M${\rm _V < -8^m}$).

\begin{figure}[ht!] 
\includegraphics[angle=0,width=0.7\textwidth,height=0.5\textheight,bb=40 50 720 530,clip]{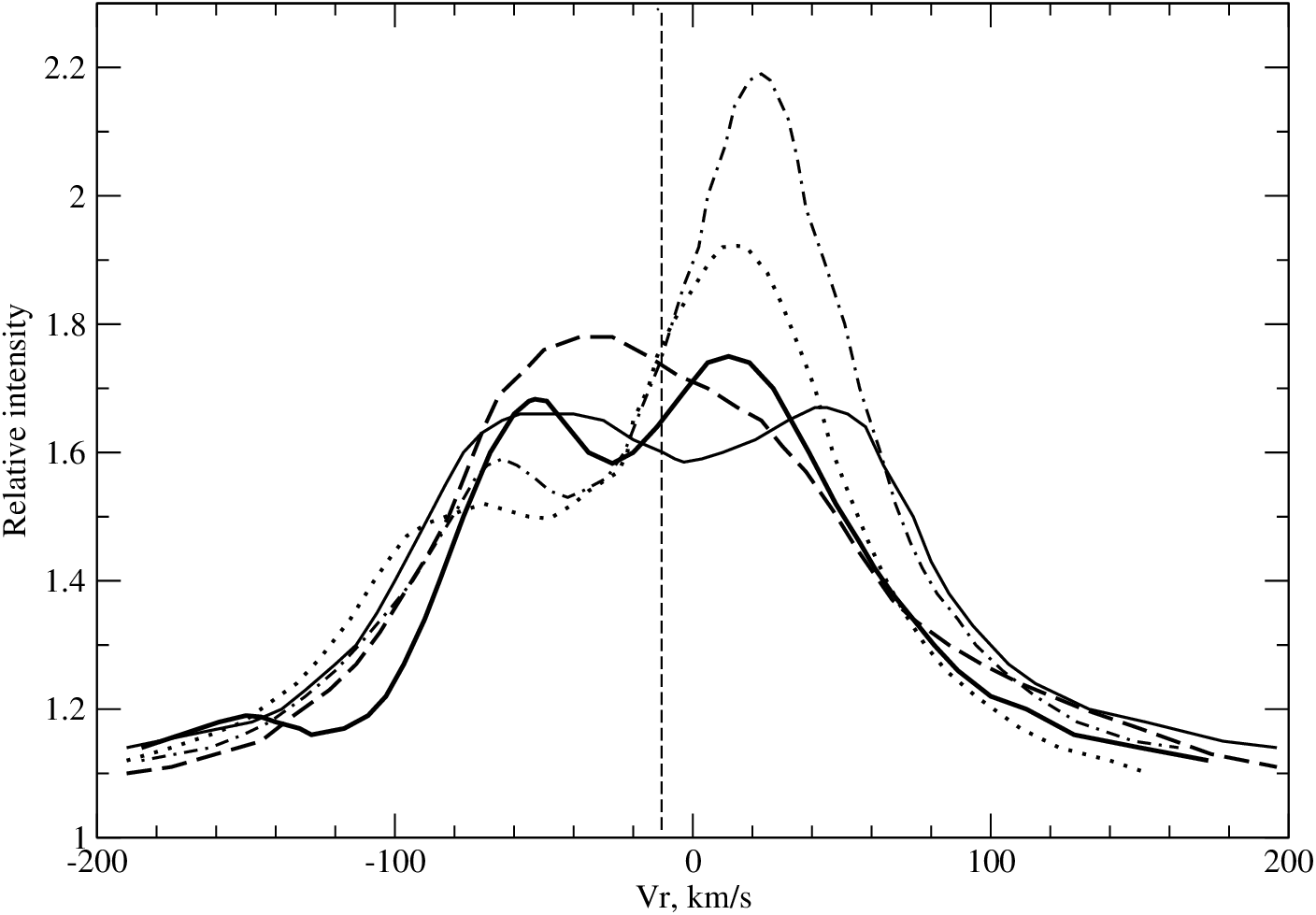}
\caption{ Variability in the H$\alpha$ profile in spectra of the hot hypergiant star No.\,12 in the Cyg\,OB2 
   association obtained on the 6-m telescope during 2001--2011 [12, 20]. The position of the vertical 
   dashed line corresponds to the systemic velocity,   Vsys\,=$-10.5$\,km/s~[20].}
\end{figure}      

Only a few such objects are known in our Galaxy, including members of stellar associations: 
Cyg~OB2 No.\,12 (B5\,Ia-0)~[12], HD\,168625 (B6\,Ia-0), and HD\,168607 (B9\,Ia-0) in the 
Ser\,OB1A association~[19]. Star No.\,12 in the Cyg\,OB2 association is the best example 
of a hot hypergiant; its membership in this association implies the reliable luminosity
$\log (L/L_{\odot})$=\,6.28~[11]. The high luminosity of the star is also indicated by 
the high intensity of the OI~7773\,\AA{}  triplet. Its equivalent width is W$_{\lambda}$\,=\,1.14\,\AA{}~[12], 
which corresponds to the absolute magnitude M${\rm _V < -8^{m}}$. This is one of the brightest
stars in the Galaxy, and has the very high initial mass  $\mathcal{M}=110\mathcal{M}_\odot$~[11].

The properties of the stellar wind of Cyg\,OB2 No.\,12 are clearly manifest in Fig.\,1, which presents
the H$\alpha$ emission profiles in the spectra of this hypergiant obtained on the 6-m telescope at 
various epochs in the interval 2001$\div$2011~[12, 20]. The vertical dashed line indicates 
the velocity of the center of mass of the system (which we will refer to as the systemic velocity), 
Vsys\,=$-10.5$km/s~[20]. The diversity of these  profiles reflects the instability of the hypergiant wind. 
Figure\,1 shows that the profile shape is variable, but its main properties are preserved:
powerful emission with a dip on the short-wavelength side, a jagged peak, and extended 
Thomson wings.
The limiting wind speed is about 150\,km/s. The inversion of the intensity in the upper part of the H$\alpha$
profile testifies that the wind of Cyg\,OB2 No.\,12 is inhomogeneous. In addition to the fast component
noted above, the wind contains a fairly large amount of material that is nearly stationary relative to the star,
or even falling onto the star. The coexistence in a single spectrum of lines with P\,Cygni and inverse P\,Cygni 
profiles, and even a combination of the two in a single line profile, has been noted for some LBVs
at the phase of their maximum brightness~[21]. This coexistence leads us to reject spherically symmetrical
models for the wind.

Let us compare the H$\alpha$ profile in the spectrum of star No.\,12 in Cyg\,OB2 with the corresponding 
profiles in the spectra of supergiants of various masses. Figure~2 presents H$\alpha$ profiles 
in the spectra of the stars we have chosen for comparison: the classical supergiants $\alpha$\,Per 
and HD\,102878, the A hypergiants HD\,33579 and 6\,Cas, and the post-AGB supergiants
V510\,Pup and BD+48\degr1220. For convenience in comparing the profiles for the various stars, 
the horizontal axis plots the shift $\Delta$Vr relative to the systemic velocity for each object. 
As expected, the H$\alpha$ profiles in the spectra of the A hypergiants HD\,33579 and 6\,Cas 
are similar to the H$\alpha$ profile of star No.\,12. However, the emission-absorption profiles 
of this line in the spectra of the post-AGB supergiants V510\,Pup and BD+48\degr1220 contain 
equally powerful emission. They are similar to the observed profiles in the spectra of hypergiants, 
but the velocity of the matter outflows are much lower in post-AGB stars. At the same time, 
the profiles of Population~I supergiants (dashed curves in Figs.\,2 and 3) differ sharply from both of
these.

\begin{figure}[ht!]      
\includegraphics[angle=0,width=0.6\textwidth,height=0.8\textwidth,bb=57 50 530 710,clip]{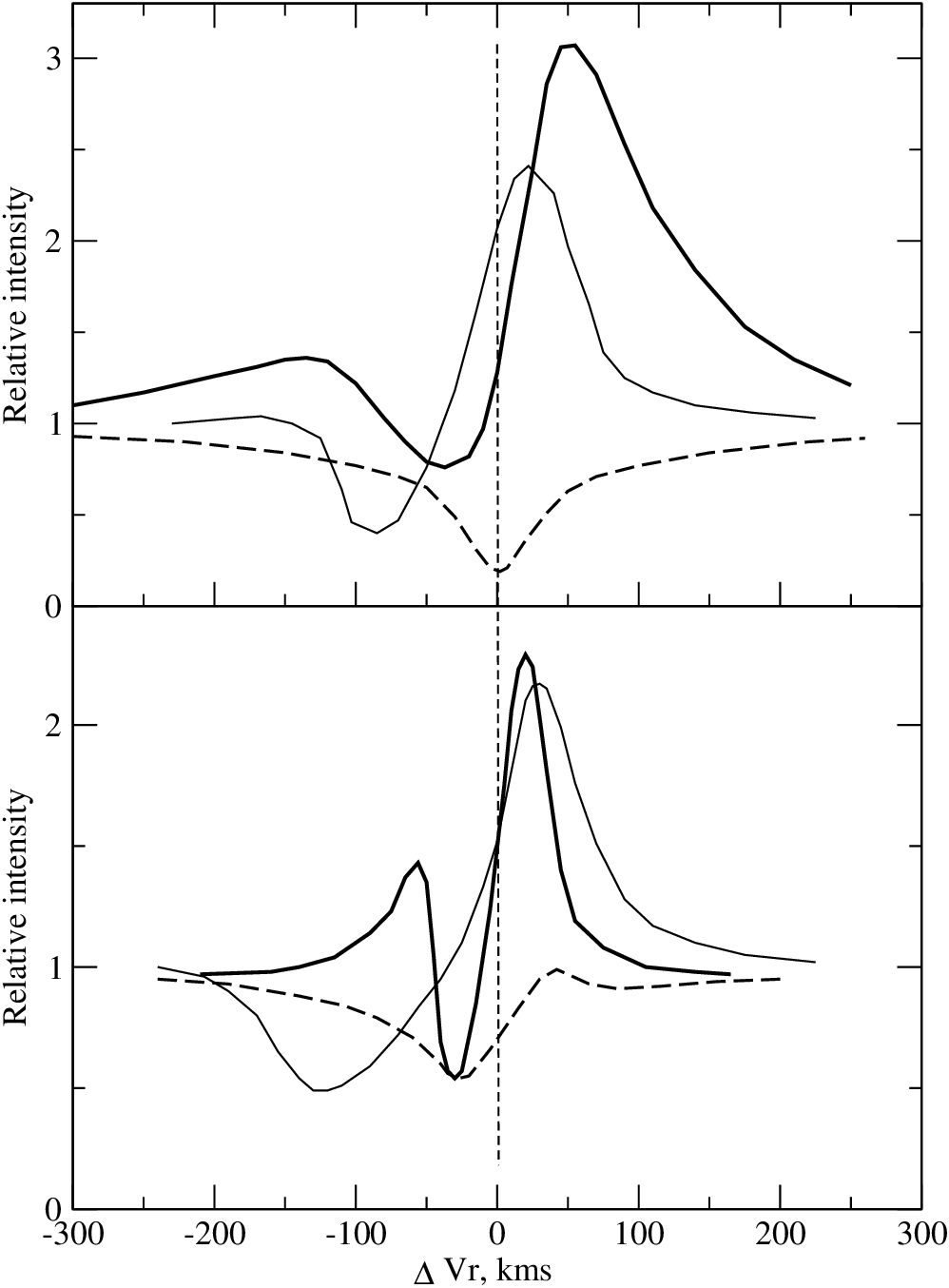} 
\caption{H$\alpha$ profiles in the spectra of supergiants. The upper panel shows profiles for $\alpha$Per 
   (dashed), HD\,33579 (thin), and V510\,Pup (bold). The lower panel shows profiles for HD\,102878 (dashed), 
   6\,Cas (thin), and BD+48\degr1220 (bold). The vertical dashed line represents the systemic velocity of 
   each object, and the horitonzal axis plots the shift $\Delta$Vr relative to the systemic velocity.}
\end{figure}

Given the well studied nature of the hypergiant 6\,Cas and the fact that its spectral type is later than
those of the supergiants listed above, we used this star in our comparisons with the spectra of A stars
with other masses. Figure~2 shows that the longwavelength H$\alpha$ absorption in 6\,Cas makes a 
transition to strong emission, forming a P\,Cygni type profile, which is a sign of a spherically 
symmetrical wind. The mean velocity indicated by the broad absorption component is close 
to Vr\,=$-150$\,km/s. The picture of a stable and spherically symmetrical wind is acceptable for 
white hypergiants in a first approximation,  although with various caveats. For example, the
main background component of the wind giving rise to the broad absorption features in the H$\alpha$ 
P\,Cygni type profile in the spectrum of 6\,Cas, which are deformed by moving dips, and the nearly 
symmetrical bell-like H$\alpha$ profile in Cyg\,OB2 No.\,12 (Fig.\,1). Additional information is 
provided by Fig.\,3, which presents H$\alpha$, H$\beta$, and H$\gamma$ profiles in the spectra 
of stars with various luminosities.

\begin{figure}[t]      
\includegraphics[angle=0,width=0.5\textwidth,height=0.8\textwidth,bb=40 50 510 755,clip]{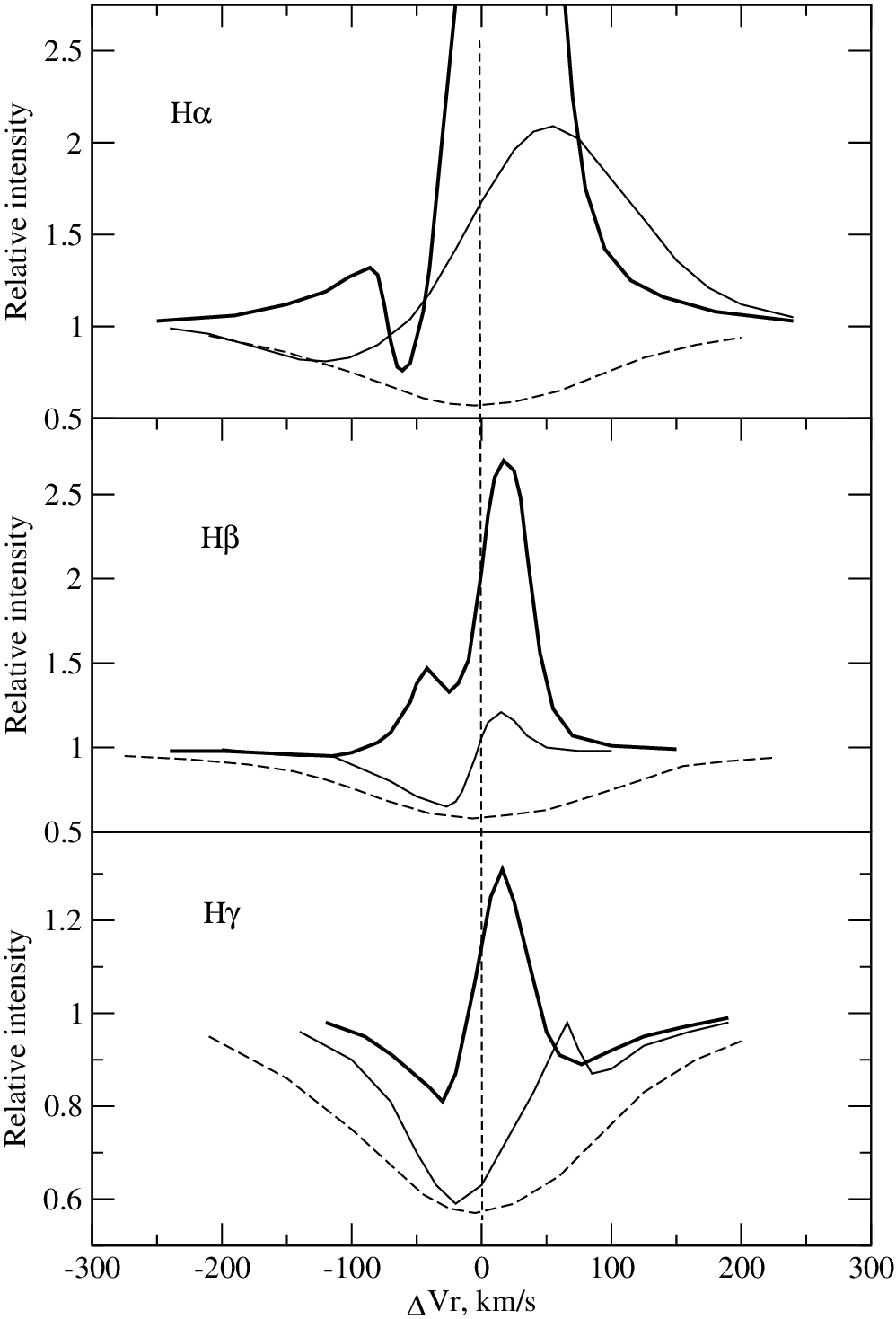} 
\caption{Neutral-hydrogen line profiles in the spectra of supergiants. The upper, middle and lower panels show
         profiles of H$\alpha$, H$\beta$, and H$\gamma$, respectively. Results are shown for HD\,168571 (dashed), 
         HD\,190603 (thin), and GSC\,04501$-$00166 (bold). The horizontal axis plots the shift $\Delta$Vr 
         relative to the systemic velocity of each object. }
\end{figure}

While they appear to be analogs of post-AGB stars according to a number of their observational properties 
(high luminosity, peculiar spectral features,  the presence of circumstellar molecular envelopes, etc.), 
yellow hypergiants have a fundamentally different nature. Their progenitors are massive (with initial 
masses  $\ge20\mathcal{M}_\odot$) and have very high absolute magnitudes;
after leaving the main sequence, they lose a substantial fraction of their mass and fall into the region of
red supergiants, then into the region of yellow hypergiants. The typical luminosity of a yellow hypergiant
is $\log(\rm L/L_\odot)\approx 5.3\div5.9$~[22]. On the Hertzsprung--Russell diagram, these objects are 
located near the highest luminosities, in the region of instability containing hypergiants with spectral 
types from A to M~[4, 23]. The structured circumstellar envelopes of hypergiants are formed through 
several episodes of strengthening of the powerful stellar wind (with mass-loss rates of 
$10^{-4}\div10^{-3}\mathcal{M}_\odot/$yr), and, as a rule, are powerful sources of IR and masers emission, as
well as emission in various molecular lines.

Studies of hypergiants and searches for hypergiants in other galaxies are of special interest because it 
is precisely such stars that are believed to be the progenitors of massive SNe\,II and SNe\,Ibc supernovae. 
The main steps in fixing the status of a hypergiant are determining its luminosity and mass loss parameters, 
identifying characteristic features in its optical spectral energy distribution, deriving
the properties of its atmospheric abundances, and tracing variations in its parameters.
This approach was applied to the known peculiar object IRC+10420, which is identified in the optical with
the star V1302\,Aql. Only after the discovery of a high nitrogen excess in its atmosphere in the late
1990s~[10], together with the detection of dramatic time variations of its parameters, was this object
finally classified as a yellow hypergiant~[10, 24, 25, 27, 28]. Moreover, V1302\,Aql has in recent years
come to be considered the most convincing example of a yellow hypergiant, and its collected observational
properties serve as a sort of template in studies of stars considered to be candidate yellow hypergiants
(see, e.g.,~[29, 30]).

Distinguishing characteristics of the optical spectra of hypergiants are powerful emission (often twopeaked) 
in the Balmer and Paschen lines of hydrogen, which have broad wings due to Thomson scattering,
and the presence of forbidden emission lines of metals. However, these two features are also present in the
spectra of high-luminosity B[e] stars. The properties of these massive, hot supergiants can be found in~[31,
32], which present good examples of optical spectra obtained with high spectral resolution. The spectra
of B[e] stars are characterized by emission in lines of the CaII IR triplet and in the Ca[II]\,7291 and 7324\,\AA{} 
forbidden lines. These profiles are in good agreement with the hypothesis that a rotating circumstellar disk
is present in a B[e] star system. The atlas~[33], which compares the spectra of the B[e] star MWC\,314 and
the hypergiant V1302\,Aql, is also useful for more detailed studies of B[e]-star spectra. We also 
recommend the study of Aret et al.~[34], which contains a compilation of the observed properties of 
a sample of B[e] stars.

Thus, we must bear in mind that the combination of intense (often two-peaked) HI emission and
forbidden emission lines of metals on its own does not provide a firm basis for classifying a star as a
hypergiant. Spectral evidence that a star is a hypergiant is provided by the combination of 
the absorption spectrum of a supergiant with powerful HI emission and emission in 
forbidden lines of metals. A high luminosity, high mass-loss rate with a modest outflow speed, 
and a detailed analysis of the pattern of the radial velocities measured from lines with different 
origins are crucial when classifying a star as a hypergiant. This problem is illustrated clearly
in the spectral atlas~[33], which compares spectra of the B[e] star MWC\,314 (Sp\,=\,B3\,Ibe) 
and the hypergiant V1302\,Aql (Sp\,=\,F5\,Ia$^+$e at the epoch used to compile the atlas). 
These stars have different masses, luminosities, and spectral types, but have similar spectra. 
In particular, the powerful hydrogen emission lines have two-peaked profiles. The spectra both 
contain numerous permitted and forbidden emission lines of metals and features with P\,Cygni
profiles.

Stars on the asymptotic giant branch (AGB) have low-mass cores with typical masses $\approx0.6\mathcal{M}_\odot$,
surrounded by an extended and often structured gas-dust envelope, which is formed via the loss of a
substantial fraction of the star’s mass in previous stages of its evolution. AGB stars make the 
transition to the planetary nebula phase with essentially constant luminosity, while becoming increasingly
hotter. They have the observational characteristics of nebulae, and are often named  as ``protoplanetary
nebulae'' (PPN). One reason for the interest of both observers and theoreticians in AGB and post-AGB
stars is that nucleosynthesis and the subsequent transport of carbon and heavy elements synthesized
in the s-process to the stellar surface (third mixing) occur in these evolutionary stages~[35]. 
As a result, intermediate-mass stars are the main suppliers to the interstellar medium of heavy 
elements (Sr, Y, Ba, La, and others), and also of a substantial fraction of C and N~[36], which 
affect the chemical evolution of the galaxy.

In spite of the differences in their masses, internal structures, and ages, intermediate-mass stars (with
initial masses of $3\div8 \mathcal{M}_\odot$) in advanced stages of their evolution have spectra similar 
to those of true (i.e., massive) hypergiants. Moreover, we can also see an analogy in the structures of 
hypergiants and post-AGB stars: in both cases, the object has an evolved core surrounded by a gas-dust 
envelope formed in prior stages of its evolution. Post-AGB stars cannot immediately be distinguished 
from supergiants or hypergiants based only on optical spectroscopy. These are most often F--G stars, 
although objects with earlier spectral types are also encountered. The high luminosities of these objects 
correspond to MK luminosity classes Ia,b, and II. However, this only indicates a low
surface gravity in the stellar photosphere (or pseudo-photosphere). Post-AGB stars are often located at
high Galactic latitudes, which is not characteristic of massive stars. The main types of optical 
spectral features displayed by PPN are: 
\begin{enumerate}
 \item{metallic absorption lines with low or moderate intensities,  symmetrical profiles without visible distortions;} 
 \item{time variable neutral-hydrogen absorption and  emission components;} 
 \item{strong metallic absorption lines with low excitation potentials for their  lower levels, with their 
 variable profiles often distorted by features arising in the envelope;} 
 \item{molecular absorption or emission bands, primarily for molecules containing carbon;}
 \item{envelope components of the NaI, KI resonance lines;} 
 \item{narrow permitted or forbidden emission lines of  metals formed in the envelope. }
\end{enumerate}
The presence of features 2$\div$6 represents the main differences  between the spectra of 
PPN and of classical massive supergiants~[37]. It is clear that all this variety of spectral features 
is lost in spectra with low resolution.

The H$\alpha$ lines in PPN spectra display complex (a combination of emission and absorption components), 
time-variable profiles of various types: with an asymmetric core, normal or inverse P~Cygni type profile,
with two emission components in the wings. Combinations of similar features are also observed fairly
often. The presence of H$\alpha$ emission indicates a high mass-loss rate and is one criterion in searches 
for PPN. The mass-loss rates of AGB stars are two to three orders of magnitude lower than those of hypergiants. 
The maximum mass-loss rate, rarely achieved in the AGB stage, is $10^{-5}\mathcal{M}_\odot$/yr (see, e.g., the data
of~[38]). In addition, the outflow velocities do not exceed about 10$\div$20\,km/s --- an order of magnitude
lower than the outflow velocities of hypergiants. Only in the so-called ``superwind'' phase (with a duration
of several hundred years) in the transition from a spherically symmetrical wind in the AGB stage to
a collimated outflow in the post-AGB stage does the mass-loss rate grow by an order of magnitude
(see~[39] and references therein).

The complex character of the variability of profiles of features in PPN spectra is due to the fact that 
binary and pulsational instability can be superposed on variability due to inhomogeneity of the wind. 
A compilation of data on the periods and types of pulsation is presented in~[40]. The main properties of 
the optical spectra of PPN are clearly illustrated by the results of studies of the spectral variability 
of HD 56126~[41], and by the atlas [42] of its spectra. According to its collected observational 
properties (a typical twopeaked spectral energy distribution, the spectrum of an F supergiant with 
variable H$\alpha$ absorption and emission profiles, the presence of Swan bands of the
C$_2$ molecule in the optical, which form in the outflowing extended envelope, high excesses of carbon
and heavy metals synthesized in the s-processes and  dredge-upped to the surface by mixing), 
this star can be considered as a canonical post-AGB star. The degree of variability of its H$\alpha$ profile 
is surprising: as follows from Fig.\,2 in the atlas~[42], all the profile types
listed above were encountered in observations with the SAO 6-m telescope obtained over about a decade:
an asymmetric core, normal or inverse P~Cygni type profile, two emission features in the wings.

The weak central stars of the PPN we are considering here  have substantially different spectral types: 
GSC\,04501$-$00166 (=IRAS\,01005+7910) has spectral type B1.5\,Ib~[14], BD+48\degr1220 (=IRAS\,05040+4820) 
has type A4\,Ib~[16], and V510\,Pup (IRAS\,08005$-$2356) has type F5\,Ib-Iab with emission lines~[17]. 
The H$\alpha$ profiles of all three stars have similar shapes. Their emission-absorption
profiles are P\,Cygni in type, with the emission component appreciably exceeding the continuum level.
An important property of the H$\alpha$ profiles in the spectra of these PPN is that the relative 
intensity of the H$\alpha$ emission is close to, or even exceeds, the corresponding relative intensities 
in the spectra of hypergiants with similar spectral types. Of the post-AGB stars presented in the Table, 
IRAS\,01005+7910 provides the most striking example of spectral mimicry of a massive hypergiant by a 
low-mass star in the post-AGB stage.  However, as follows from Figs.\,2 and 3, the velocities of wind 
absorption  features in post-AGB supergiants are a factor of two to three lower than those in massive stars. 
Such relatively low wind speeds are characteristic of hypergiants, rather than classical supergiants.

The profiles of metal lines in the spectra of PPN also differ sharply from the norm; this is clearly visible
in Fig.\,4, which compares profiles of the HeI~5876\,\AA{} and SiII~6347\,\AA{} lines in the spectra of the 
hypergiant HD\,190603, the classical supergiant HD\,168571, and the post-AGB star GSC\,04501$-$00166. The 
relative intensities of the emission components of both lines in the spectrum of the post-AGB star appreciably
exceed their intensity in the spectra of the both more massive supergiants. Similar behavior can be seen 
in Fig.\,5, which presents fragments of the spectra of the massive F supergiant HD\,74180 and the post-AGB star
V510\,Pup (IRAS\,08005$-$2356), which have similar spectral types.

\begin{figure}[ht!]      
\includegraphics[angle=0,width=0.8\textwidth,height=0.5\textwidth,bb=40 50 720 530,clip]{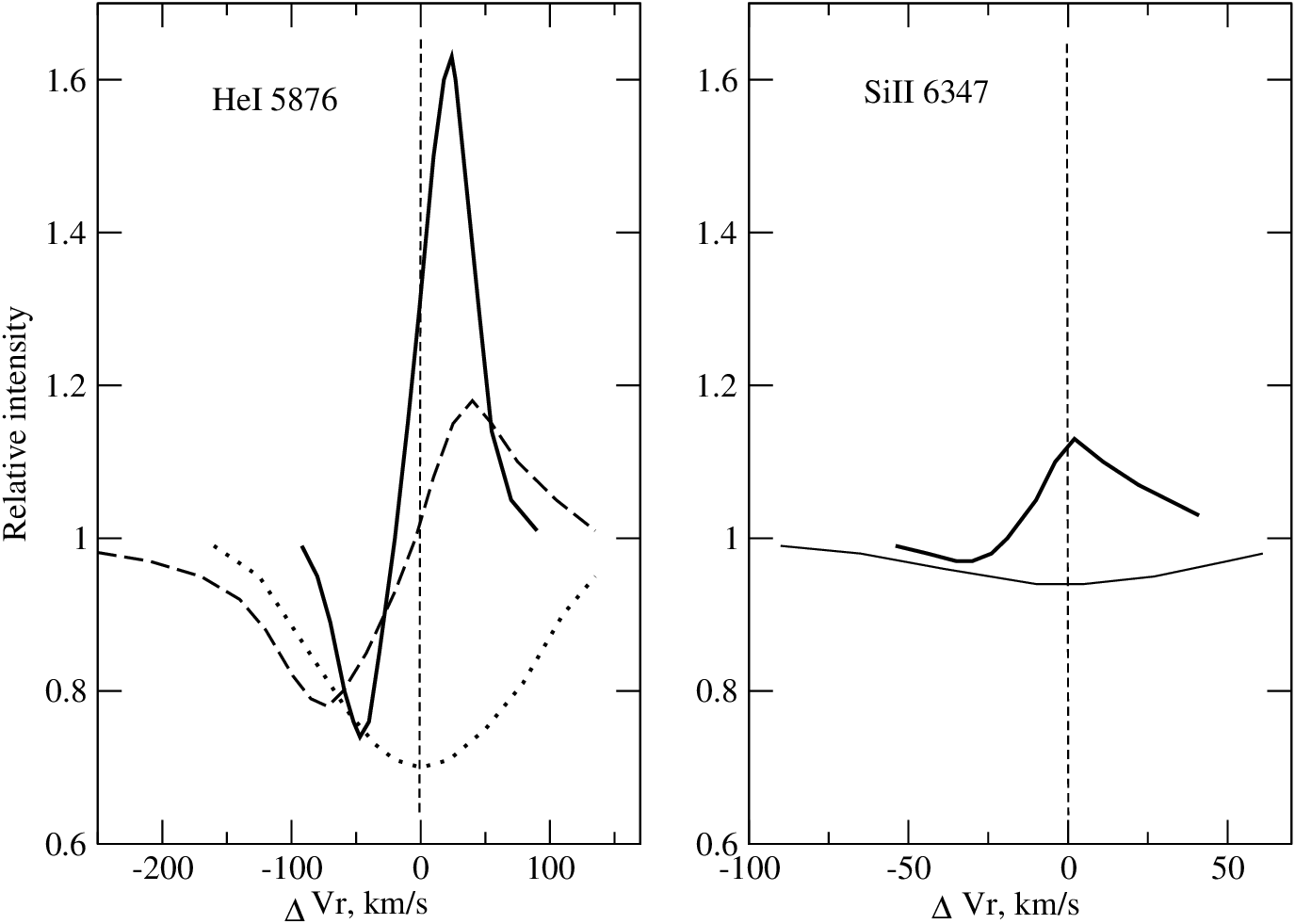} 
\caption{Profiles of selected lines in the spectra of supergiants: HeI\,5876\,\aa{} is shown to the left 
       (the dotted profile is HD\,168571, dashed is HD\,190603, and bold is GSC\,04501$-$00166), and 
       SiII~6347\,\aa{} is shown to the right (the thin profile is HD\,148379 and bold is GSC\,04501$-$00166). 
       The horizontal axis plots the shift $\Delta$Vr relative to the systemic velocity of each object.}
\end{figure}

\begin{figure}[ht!]      
\includegraphics[angle=0,width=0.7\textwidth,height=0.4\textwidth,bb=40 50 720 530,clip]{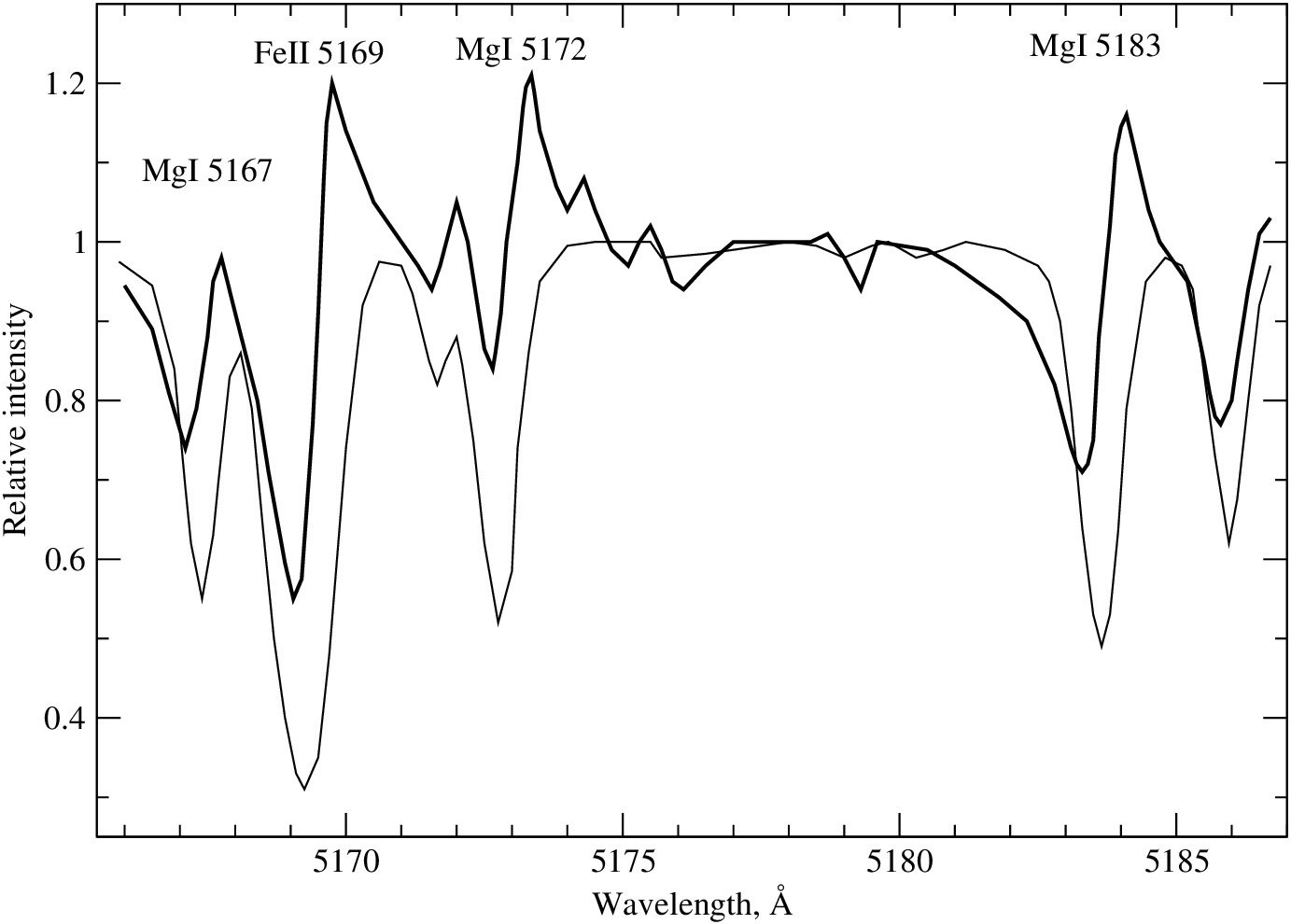} 
\caption{Comparison of fragments of the spectra of V510\,Pup (bold) and HD\,74180 (F0\,Ia) (thin) at wavelengths 
          $\lambda\lambda 5165\div5187$\,\AA{}. The horizontal axis plots the laboratory wavelength.}
\end{figure}

In relation to the problem of mimicry, it is also useful to consider the observational properties of
the poorly studied variable V2324\,Cyg (it has visible magnitude V\,=\,11.63$^m$, color indices 
B-V\,=+1.09$^m$ and U-B\,=+0.58$^m$, and Galactic coordinates l\,=\,89.44\degr, b\,=+2.39\degr, 
and the optical object is identified with the IR source IRAS\,20572+4919).
The observed excess at 12$\div60 \mu$ and its position on an IR color diagram suggests 
that this object may be a young planetary nebula with a dust envelope~[43].
Little is currently known about the properties of V2324\,Cyg. Arkhipova et al.~[44, 45] detected
variability in the brightness of V2324\,Cyg in longtime UBV observations, with amplitudes $\approx0.3^m$ 
in U and $\approx 0.2^m$ in V and B. They explained the brightness variability as being due to the influence 
of the stellar wind. No pulsational periodicity was detected in the brightness variability~[45]. 
The absence of pulsations is consistent with the fairly early spectral type of the star, A3\,I~[45]. 
Garcia-Lario\,et\,al.~[46] used IR photometric data to determine the evolutionary statuses of 
225 IRAS sources; they consider the classification of IRAS\,20572+4919 as a post-AGB star to be 
tentative. A detailed study of the optical spectrum of V2324\,Cyg was carried out in~[18],
where it is shown that V2324\,Cyg has an atypically high rotational velocity for a post-AGB star, 
$V\sin i$\,=\,69\,km/s. The spectral type F2\,III was determined for V2324\,Cyg based on metallic 
lines of moderate intensity. The lower limit of the luminosity~[18] leads to a contradiction with the 
properties of the chemical composition of its atmosphere. According to current thinking (see, e.g., the 
review~[47]), large excesses of lithium and sodium provide evidence that the star is a
post-AGB object which had an initial mass exceeding 4\,$\mathcal{M}_\odot$.

\begin{figure}[ht!]      
\includegraphics[angle=0,width=0.7\textwidth,height=0.6\textwidth,bb=40 50 720 530,clip]{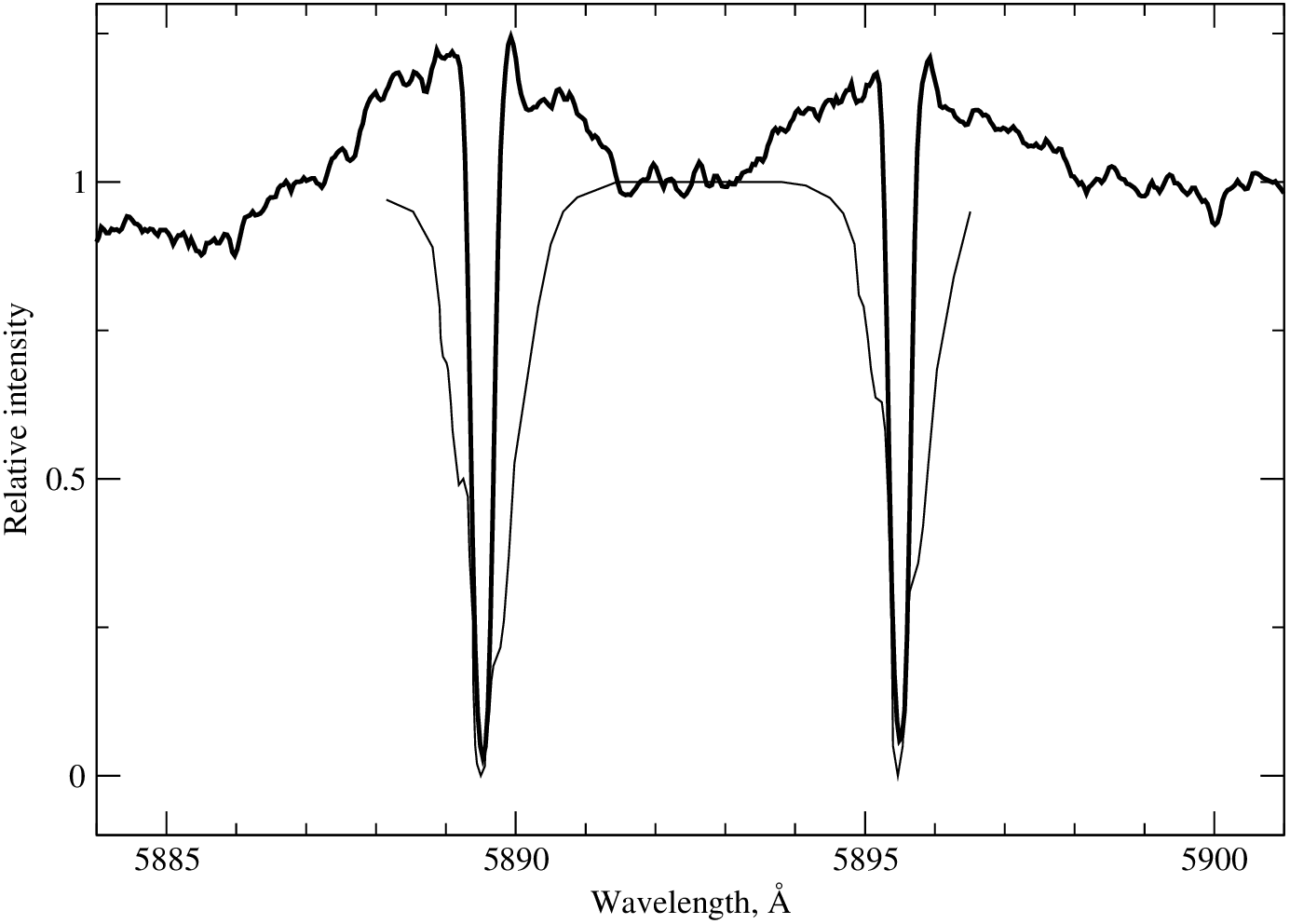} 
\caption{Comparison of profiles of the NaI~D lines in the spectra of V2324\,Cyg (F2\,III) (bold) and 
        HD\,74180 (F0\,Ia) (thin). The horizontal axis plots laboratory wavelength.}
\end{figure}

\begin{figure}[ht!]      
\includegraphics[angle=0,width=0.7\textwidth,height=0.6\textwidth,bb=40 50 720 530,clip]{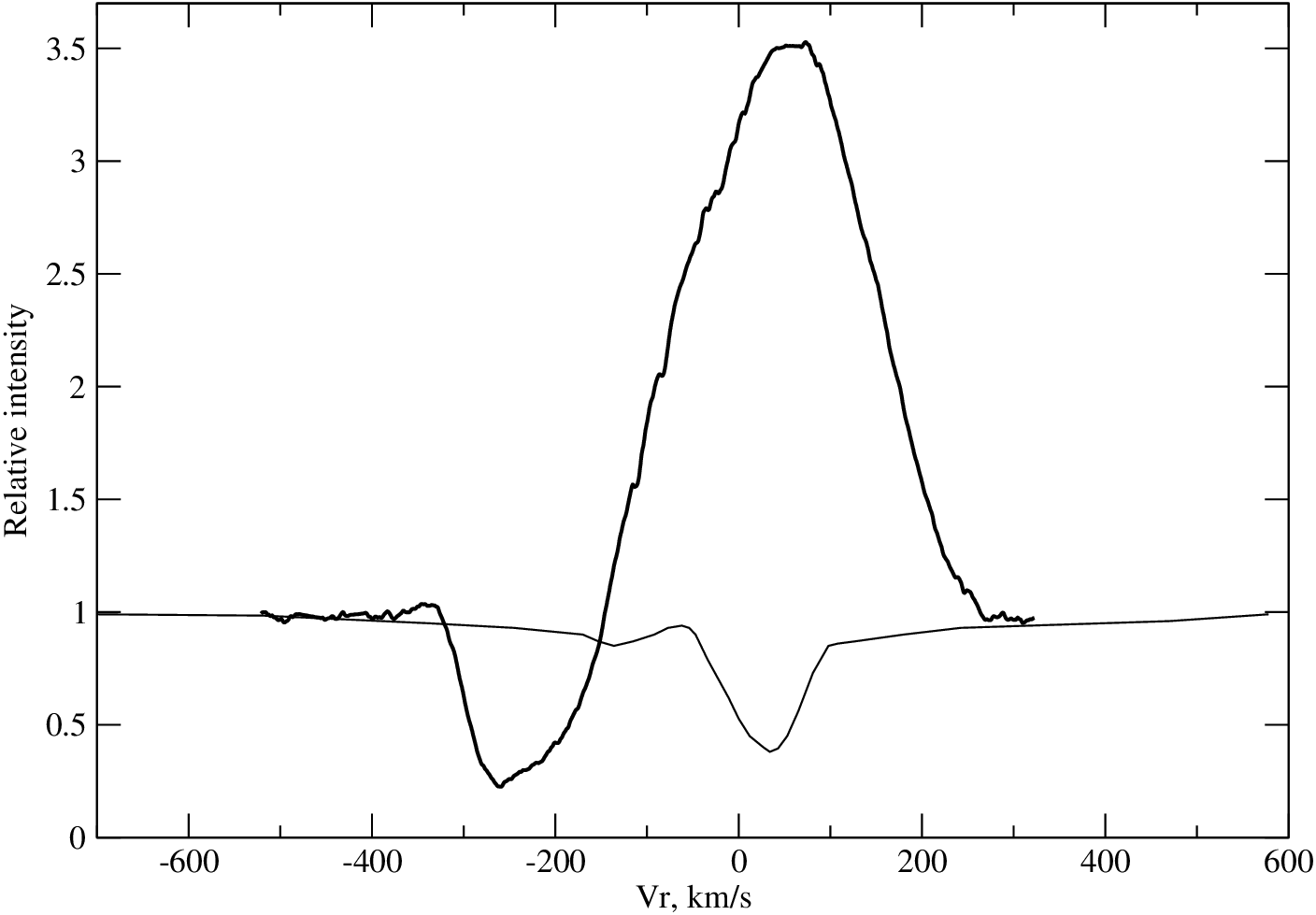} 
\caption{H$\alpha$ profile in the spectra of V2324\,Cyg (F2\,III) (bold) and HD\,74180 (F0\,Ia) (thin) } 
\end{figure}

In connection with the publishing of the Gaia data, we are now able to refine the luminosity of
V2324\,Cyg. Taking into account interstellar reddening, E(B$-$V)\,=\,1.1$^m$~[44], this object\,'s 
parallax measured by Gaia, $\pi$\,=\,1.55\,mas, implies a low absolute magnitude, M$_{\rm V}$\,=$-1.0^m$. 
However, some  features of its spectrum sharply distinguish V2324\,Cyg from ordinary early-subclass 
F~giants, most importantly the presence of emission in a number of lines. As can be seen in Figs.\,6 and 7, 
emission is present in the NaI\,(1) doublet, and is especially strong in H$\alpha$ (the red component 
of its P\,Cygni type profile). The H$\alpha$ emission peak exceeds the continuum level by a
factor of three to four. A comparison of the H$\alpha$ profiles in spectra obtained at different 
epochs indicates variability: the profile shape, emission intensity, and absorption depth all vary~[18]. 
The positions of the emission peak and of absorption lines also vary. Overall, doubt still remains 
as to whether this star is in the post-AGB stage.

\section{Main results and conclusions} 

The mass-loss rate grows with a star`s luminosity, due to increasing density of the wind, since its
speed is reduced. For example, according to~[48], these parameters for $\alpha$\,Cyg (A2\,Ia), the very 
bright LMC white hypergiant HD\,33579 (A3\,Ia-0), and S\,Dor at their maximum brightnesses are related as
follows: the mass-loss rate grows from the first to the last object as 1\,:\,100\,:\,1000, while the 
mean outflow velocities (based on rocket-UV resonance lines) are 200, 120, and 70\,km/s, respectively. 
This type of correlation is also obtained for our objects. Comparing the limiting outflow velocities 
derived from the H$\alpha$ absorption lines and the mass-loss rates (from data in the literature), 
we conclude that the winds of supergiants and hypergiants are aspherical and nonstationary, with 
their spatial and kinematic structures simplifying with increasing luminosity [49]. Note that
the question of spectral mimicry is one aspect of the general problem of fixing the evolutionary 
status of evolved stars with various masses. We note here the recent study of Humphreys~et\,al.~[50], 
who considered spectral mimicry among the most massive HLSs with optical emission lines: B[e] supergiants, 
LBVs, LBV candidates, and Wolf-Rayet stars.

A promising direction for studies of the final stages of the evolution of stars with various initial 
masses is joint studies and analyses of the chemical compositions and velocity fields in their 
atmospheres and envelopes. Ammonia (NH$_3$) was detected in the envelopes of the hypergiants V1302\,Aql 
and HD\,179821~[6], and the atmospheres of both these stars are enriched in nitrogen~[10, 51]. 
These results strongly suggest that both objects are descendents of massive stars. Similar results 
were also obtained for a number of post-AGB stars over the past decade. First, a subsample of 
post-AGB stars was distinguished, whose atmospheres display large excesses of carbon and heavy metals, 
and whose circumstellar envelopes have complex morphologies and are usually enriched in carbon, 
manifest through the presence of bands of C$_2$, C$_3$, CN, CO and other carboncontaining molecules 
in the IR, radio, and optical spectra (see~[37] and references therein). Second, several objects whose 
atmospheres are enriched in heavy metals are found among these post-AGB stars, for which high-resolution 
optical spectroscopy also indicates the presence of heavy metals in their circumstellar envelopes~[37, 52--54]. 
We again emphasize that these spectral features cannot be studied in low-resolution, or even medium-resolution, 
spectra.

As a rule, fixing the evolutionary status of peculiar HLSs requires the use of a varied set of modern
observational methods. This is clearly demonstrated  by the history of studies of the peculiar supergiant V1302\,Aql 
(IRC+10420), which was over several decades, right up to recent years, considered to be a post-AGB star. 
Only the collected results obtained over many years of observations using various methods on the largest 
optical and radio telescopes helped identify its true nature. In addition to high-resolution optical spectroscopy, 
decisive results were obtained from spectroscopy with high spatial resolution~[24], IR spectroscopy~[55], 
and spectropolarimetry~[56]. Through these efforts, the object was finally classified
as a massive star with extremely high luminosity --- a yellow hypergiant. An example of an object with an
uncertain status is the rather poorly studied variable A~supergiant V2324\,Cyg, identified with the IR source
IRAS\,20572+4919. The collected information that is available is currently insuffcient to unambiguously
classify this as a massive or low-mass supergiant. 

The reasons the problem of spectral mimicry are caused to the fact that essentially all features in the
optical spectra of HLSs form in various layers of the extended and expanding stellar atmosphere. In recent 
studies analyzing spectra of very evolved stars of various spectral types with mass loss (see, 
e.g.,~[10, 57--60]), this extended atmosphere that is optically thick in the continuum and is formed 
by the stellar wind has sometimes been called a ``pseudo-photosphere'', following the introduction of 
this term in~[61]. The presence of these specific layers is characteristic of both hypergiants and post-AGB 
stars. The total radiation flux determining the effective temperature forms in inner, energetically 
active layers of the star, but it is difficult to determine from spectral observations which 
energy-release processes give rise to the stellar radiation with this effective temperature:
burning of hydrogen or helium in the core or in shell sources.  Apart from their similar effective  
temperatures, these processes affect appreciably different luminosities. Another important factor 
influencing the formation of spectral features is the dynamical state of the atmosphere, in 
particular, the presence of an intense stellar wind and the passage of shocks resulting from pulsations. 
The variety of types of spectral features is determined primarily by the physical conditions in the 
extended atmosphere of the star, and depends less on the mass of lower-lying photospheric
layers. This is clearly demonstrated by the presence of powerful HI emission lines in the spectra 
of post-AGB stars, whose relative intensities often exceed the corresponding relative 
intensities in the spectra of hypergiants.

In conclusion, we will focus on the known fact that HLSs are widely used as standard candles 
to construct  a distance scale both  in our Galaxy and extending to extragalactic distances. 
The spectral mimicry of supergiants  must especially be borne in mind when using HLSs to 
determine the distances of distant objects. 
Unreliable estimates of an object`s luminosity, and consequently its distance, can be obtained 
if one relies purely on the results of spectral classification. The luminosity of a star can be
determined both from its membership in some group at a known distance, or from the dynamical 
properties of its atmosphere. The former method is rarely used, since stars located in 
short-lived evolutionary stages are very rarely encountered in clusters and associations. 
When applying the latter method, we often encounter the problem of spectral mimicry, when the
same type of spectrum can be associated with objects with very different evolutionary statuses. 
The evolutionary status can be refined based on the changed chemical composition of the 
atmosphere (if nucleosynthesis products have already been transported to it). Few such 
cases are known (they comprise 6–7\% of the sample studied in~[54]). A second method is
long-term spectroscopic monitoring, which can reveal properties of the pseudo-photosphere 
that cannot be determined from a single spectrogram.

As we have already noted above, the most important, but also the most difficult, aspect of our
studies is fixing the luminosities of the stars. There is now hope that this problem will be 
eased even for distant objects with the accumulation and publication of astrometric data from 
the Gaia mission. Experience working with numerous and varied stellar spectra suggests another 
effective means of resolving this problem, which is unfortunately resourceintensive:  detailed 
studies of the profiles of interstellar lines (NaI, DIBs, etc.) in the spectra of neighboring
stars as a function of their distance in the Galaxy. This approach was successfully applied 
in studies of the anomalously red star No.\,12  -- an LBV candidate in the Cyg\,OB2 association~[62]. 
With the aim of clarifying the membership of stars in the association, classifying their spectra, 
determining their reddening, and investigating the behavior of reddening within the association, 
Maryeva et al.~[62] obtained extensive additional spectral material for probable members of
the association. A result of this study is the conclusion that the excess reddening of star 
No.\,12 arises in its circumstellar envelope, which formed via high mass loss in prior evolutionary 
phases. 

The resolution of the problem of mimicry for each individual object consists first and 
foremost in determining its evolutionary status, which requires determining and comparing 
various parameters: position in the Galaxy, luminosity, wind parameters, spectral
energy distribution, chemical composition, and a detailed picture of the velocities 
at different levels in the stellar atmosphere and in its circumstellar medium.

\acknowledgements
Observations on the SAO 6-m  telescope are supported by the Ministry of Education and 
Science of the Russian Federation (agreement No.\,14.619.21.0004, project code RFMEFI61914X0004).
This work has made use of the SIMBAD and ADS astronomical databases and data from the  ESA 
mission Gaia ({\it https://www.cosmos.esa.int/gaia}).

\end{document}